\documentclass[conference]{IEEEtran}

\IEEEoverridecommandlockouts
\usepackage{cite}
\usepackage{amsmath,amssymb,amsfonts}
\usepackage{algorithmic}
\usepackage{graphicx}
\usepackage{textcomp}
\usepackage{url}

\usepackage{multirow}
\usepackage{booktabs,makecell}

\usepackage{xcolor}
\def\BibTeX{{\rm B\kern-.05em{\sc i\kern-.025em b}\kern-.08em
    T\kern-.1667em\lower.7ex\hbox{E}\kern-.125emX}}
\begin{document}


\title{Towards Unified Music Emotion Recognition across Dimensional and Categorical Models}

\author{
\IEEEauthorblockN{Jaeyong Kang}
\IEEEauthorblockA{
\textit{Information Systems, Technology and Design} \\
\textit{Singapore University of Technology and Design}\\
Singapore \\
jaeyong\_kang@sutd.edu.sg
}
\and
\IEEEauthorblockN{Dorien Herremans}
\IEEEauthorblockA{
\textit{Information Systems, Technology and Design} \\
\textit{Singapore University of Technology and Design}\\
Singapore \\
dorien\_herremans@sutd.edu.sg
}
}


\maketitle

\begin{abstract}
One of the most significant challenges in Music Emotion Recognition (MER) comes from the fact that emotion labels can be heterogeneous across datasets with regard to the emotion representation, including categorical (e.g., happy, sad) versus dimensional labels (e.g., valence-arousal). In this paper, we present a unified multitask learning framework that combines these two types of labels and is thus able to be trained on multiple datasets. This framework uses an effective input representation that combines musical features (i.e., key and chords) and MERT embeddings. Moreover, knowledge distillation is employed to transfer the knowledge of teacher models trained on individual datasets to a student model, enhancing its ability to generalize across multiple tasks. To validate our proposed framework, we conducted extensive experiments on a variety of datasets, including MTG-Jamendo, DEAM, PMEmo, and EmoMusic. According to our experimental results, the inclusion of musical features, multitask learning, and knowledge distillation significantly enhances performance. In particular, our model outperforms the state-of-the-art models on the MTG-Jamendo dataset. Our work makes a significant contribution to MER by allowing the combination of categorical and dimensional emotion labels in one unified framework, thus enabling training across datasets.
\end{abstract}


\section{Introduction}
\label{sec:intro}

Music plays an essential role in influencing human emotions~\cite{meyer1954emotion}. In the past decades, numerous Music Emotion Recognition (MER) models been developed. MER has found applications in various domains, e.g., music recommendation systems \cite{tran2023emotion}, generative systems \cite{makris2021generating}, and music therapy \cite{agres2021music}. However, the field faces significant challenges due to the diversity of available datasets and their inconsistent labeling schemes \cite{kang2024we}.

Most MER datasets can be categorized into two types: categorical labels, which represent discrete emotions such as happy or sad, and dimensional labels, such as Russell’s circumplex model of affect \cite{russell1980circumplex}, which describes emotions on continuous scales of valence and arousal. While these datasets provide valuable resources, their heterogeneous nature complicates efforts to combine them effectively. Previous approaches \cite{jia2022music, chen2020multimodal, bour2021frequency, tan2021semi, pham2021selab, mayerl2021recognizing, rajesh2020musical, chaki2020attentive} have often focused on a single type of label, limiting their ability as they can only train on one (often small) dataset.

To address these challenges, we introduce a unified multitask learning framework that incorporates both categorical and dimensional emotion labels within a shared architecture. This approach enables us to train on multiple datasets. Multitask learning has been shown to be successful on many other tasks such as speech emotion recognition \cite{cai2021speech}, symbolic music emotion recognition \cite{qiu2022novel}, genre classification and mood detection \cite{greer2023creating}, to financial portfolio prediction \cite{ong2023constructing}. These studies demonstrate the benefits of shared representations across tasks.

To properly represent the input music, our proposed framework combines three types of features: 1) Music undERstanding model with large-scale self-supervised Training (MERT) embeddings \cite{yizhi2023mert}, 2) harmonic representations of chord progressions, and 3) musical key. The MERT embeddings are able to provide a rich representation of timbre, rhythm, and high-level musical semantics. Chord progressions and musical key features are able to encode harmonic and tonal structures, which are equally vital for understanding emotion in music.

An important  innovation of our framework is the use of knowledge distillation (KD) \cite{hinton2015distilling} to unify learning across datasets with disparate label types. Pre-trained teacher models, optimized for either categorical or dimensional labels, guide the multitask student model by providing soft target logits, which are probability distributions over the possible labels. These logits help the student model learn more effectively by transferring the teacher model’s knowledge, thereby improving generalization and performance across datasets. Additionally, the framework incorporates data augmentation strategies during feature extraction to introduce variability and mitigate overfitting, ensuring robustness across diverse audio inputs.

To validate our proposed framework, we conducted several experiments on multiple datasets, including MTG-Jamendo \cite{bogdanov2019mtg} for categorical labels and DEAM \cite{aljanaki2017developing}, PMEmo \cite{zhang2018pmemo}, and EmoMusic \cite{soleymani20131000} for dimensional labels. Our experimental results show that integration of MERT embeddings with high-level musical feature (i.e., chord progression and key) significantly enhances performance. In addition, training the network on heterogeneous datasets with multitask learning can further improve performance on single datasets, demonstrating the ability to bridge the gap between diverse label types. On the MTG-Jamendo dataset, our model achieves state-of-the-art performance, surpassing the best-performing model, lileonardo \cite{bour2021frequency}, from the MediaEval 2021 competition \cite{tovstogan2021mediaeval}, as well as more recent approaches \cite{hasumi2025music,greer2023creating}, with the highest PR-AUC of 0.1543 and ROC-AUC of 0.7810. These results highlight the effectiveness of our approach in advancing MER.

In summary, our contributions are listed as follows:

\begin{itemize}

\item We developed a unified multitask learning framework for MER that facilitates training on datasets with both categorical and dimensional emotion labels. 
\item We combined the high-level musical features with MERT embeddings for a richer input representation. 
\item We conducted several experiments and showed the effectiveness of our proposed framework. In addition, our proposed model achieves state-of-the-art performance on the MTG-Jamendo dataset \cite{bogdanov2019mtg}. 

\end{itemize}

In the rest of the paper, we first describe the related work in Section \ref{sec:related}. This is followed by a description of our proposed framework in Section \ref{sec:methods}. After that, we describe our experimental setup and its results in Section \ref{sec:exp}. Finally, Section \ref{sec:conclusion} offers conclusions from this work.

\section{Related Work}
\label{sec:related}

Below we will discuss some of the existing literature in MER research. For a more comprehensive overview of the literature, the reader is referred to \cite{kang2024we, shelke2024exploring}.

\subsection{Single-Task MER Models}
Most MER models mainly concentrate on single-task learning, which involves training models on individual datasets which use particular emotion labeling schemes, like categorical labels (e.g., happy, sad) or dimensional labels (e.g., valence-arousal). Convolutional neural network (CNN)-based methods have demonstrated strong performance in MER across various datasets. For instance, Liu et al. \cite{liu2017cnn} proposed a spectrogram-based CNN model which captures temporal and spectral features. This model has a macro F1-score of 0.472 and a micro F1-score of 0.534 on CAL500 dataset \cite{turnbull2007towards}, and a macro F1-score of 0.596 and a micro F1-score of 0.709 on CAL500exp dataset \cite{wang2014towards}, outperforming traditional approaches. Bour et al. \cite{bour2021frequency} introduced the frequency-dependent convolutions in a CNN model, achieving the highest performance at MediaEval 2021 with a PR-AUC-macro of 0.1509 and a ROC-AUC-macro of 0.7748 on MTG-Jamendo dataset \cite{bogdanov2019mtg}. Recently, Jia \cite{jia2022music} introduced a CNN-based model which combines MFCCs with residual phase features, achieving a recognition accuracy of 92.06\% on a dataset comprising 2,906 songs categorized into four emotion classes: anger, happiness, relaxation, and sadness.

Given the temporal nature of music, many researchers use recurrent neural network (RNN)-based architectures, such as Long-Short Term Memory networks (LSTMs) and Gated Recurrent Units (GRUs). For instance, Rajesh et al. \cite{rajesh2020musical} used LSTMs with MFCC features to predict emotion on the DEAM dataset \cite{aljanaki2017developing}, achieving an accuracy of 89.3\%, which is better than SVM’s 85.7\%. Chaki et al. \cite{chaki2020attentive} added attention mechanisms to LSTMs to concentrate on emotionally significant segments, resulting in \(R^2\) scores of 0.53 for valence and 0.75 for arousal on the EmoMusic dataset \cite{soleymani20131000}, surpassing the performance of the LSTM variant that lacked attention mechanisms.



Recently, Transformer-based methods have seen a rise in popularity. For instance, Suresh et al. \cite{suresh2023transformer} proposed the multimodal model with Transformer-based architecture for classifying the mood of music by leveraging the features from audio and lyrics. Their proposed model was evaluated on a subset of the MoodyLyrics dataset \cite{ccano2017moodylyrics} which contains audio and lyrics of 680 songs. Their proposed model outperforms the unimodal model with an accuracy of 77.94\%.

The absence of official train/test splits in many datasets, such as the PMEmo dataset \cite{zhang2018pmemo}, makes it difficult to compare models directly, except for the MTG-Jamendo dataset \cite{bogdanov2019mtg}, which provides official splits. This enables us to compare our model with those from MediaEval 2021 \cite{bour2021frequency,pham2021selab,tan2021semi,mayerl2021recognizing} as well as more recent methods \cite{hasumi2025music,greer2023creating,li2023mert} in our experiments. Despite these achievements, single-task models frequently face challenges in generalizing across various datasets, underscoring the need for research that integrates multiple datasets.

\subsection{MER Models with Multi-Dataset Integration}
Integrating multiple datasets is essential for improving generalization but is challenging due to inconsistencies in labeling schemes, especially between categorical and dimensional labels. Liu et al. \cite{liu2024leveraging} addressed this issue by leveraging a large language model (LLM) to align categorical labels from various datasets into a common semantic space. They used three disjoint datasets with categorical labels, which include MTG-Jamendo~\cite{bogdanov2019mtg}, CAL500~\cite{turnbull2007towards}, and Emotify \cite{aljanaki2016studying}. They showed the effectiveness of their approach by performing zero-shot inference on a new dataset.

Mazzetta et al. \cite{multi-source-mer2024} introduced a multi-source learning framework that integrates features and labels from heterogeneous datasets, including 4Q \cite{panda2018musical}, PMEmo \cite{zhang2018pmemo}, EmoMusic \cite{soleymani20131000}, and the Bi-Modal Emotion Dataset \cite{malheiro2016bi} to enhance the model’s robustness. These datasets use Russell’s Circumplex Model of Affect, and focus on the valence and arousal dimensions. While this framework effectively utilizes dimensional labels to improve robustness, it does not incorporate categorical labels, thus limiting its ability to fully leverage the diversity of available datasets \cite{kang2024we}. Developing frameworks that integrate both label types remains an open challenge, which we address in this work.

\begin{figure*}[!ht]
\centering
\includegraphics[width=6.1in]{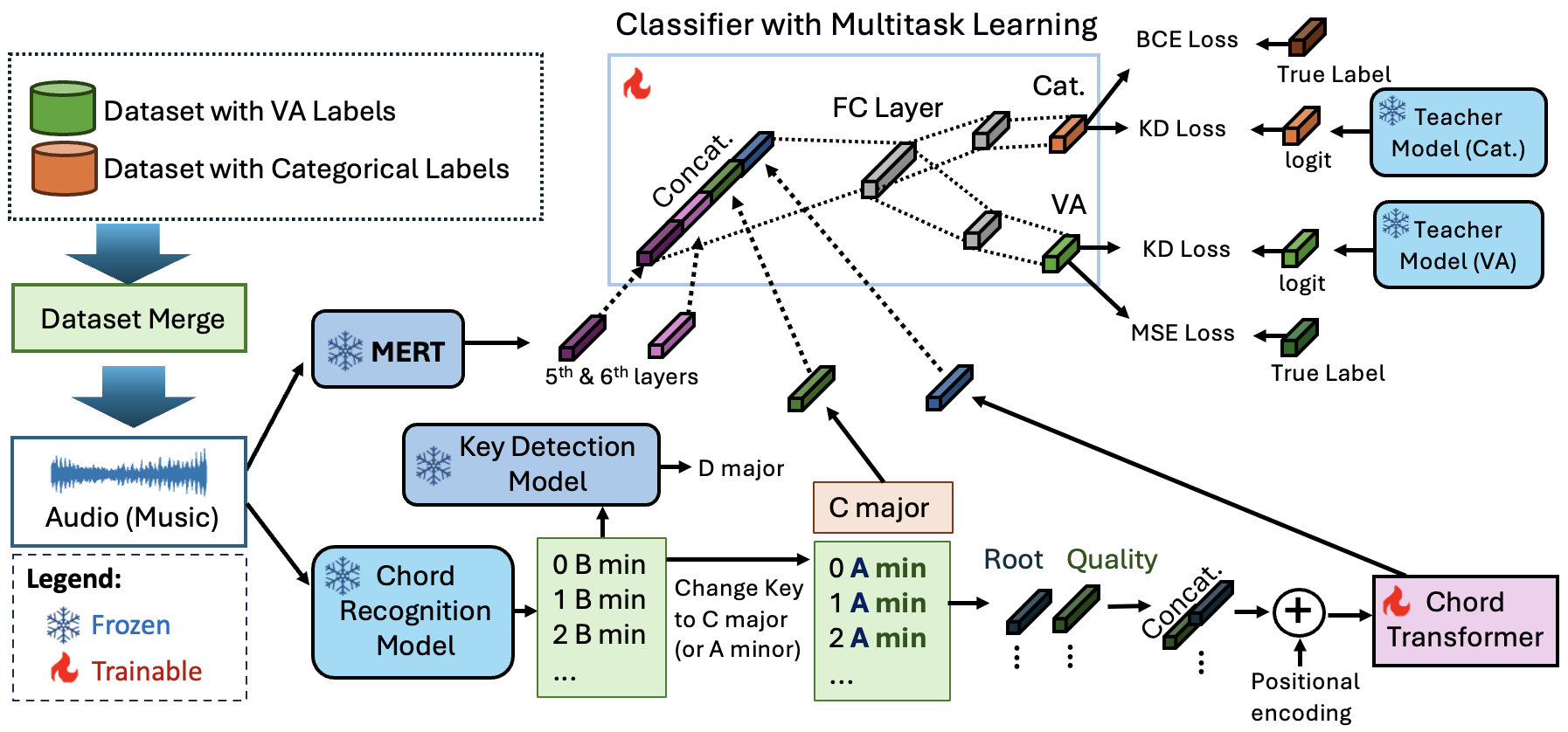}
\caption{Overall architecture of our proposed framework.}
\label{fig:framework}
\end{figure*}

\subsection{Multitask Learning in MER}
Multitask learning (MTL) \cite{caruana1997multitask} aims to enhance the performance of multiple related tasks by leveraging shared knowledge among them. Qiu et al. \cite{qiu2022novel} introduced an MTL framework for Symbolic Music Emotion Recognition (SMER). They combines emotion recognition task with auxiliary tasks such as key and velocity classification. The idea is that by forcing the model to learn key and velocity information, it will also better understand the resulting emotion. Evaluated on the EMOPIA \cite{hung2021emopia} and VGMIDI \cite{ferreira2021learning} datasets, their approach achieved accuracy improvements of 4.17\% and 1.97\%, respectively, compared to single-task models. Ji et al. \cite{huang2022adff} introduced an attention-based deep feature fusion (ADFF) method for valence-arousal prediction within a multitask learning framework. They integrate spatial feature extraction using an adapted VGGNet with a squeeze-and-excitation (SE) attention-based temporal module to enhance affect-salient features. Their model achieved \(R^2\) scores of 0.4575 for valence and 0.6394 for arousal on the PMEmo dataset \cite{zhang2018pmemo}.

Existing MTL frameworks in MER are typically still trained on datasets with one type of emotion label. In this research, we explore how we can leverage an MTL architecture to enable predicting across heterogeneous emotion datasets.

\subsection{Knowledge Distillation for MER}
Knowledge distillation (KD) is a technique where a smaller student model learns to replicate the behavior of a larger teacher model by aligning its soft predictions with those of the teacher, rather than solely relying on the actual labels \cite{hinton2015distilling}. In contrast to transfer learning, which usually involves adjusting a pre-trained model for a new task or dataset, KD emphasizes the use of the teacher model's outputs (soft labels) to guide the training of the student model.
Tong \cite{tong2022multimodal} proposed a KD-based multimodal MER framework that uses a teacher-student model, where a pre-trained genre classification model transfers knowledge to an emotion recognition model. Knowledge transfer is guided by Exponential Moving Average (EMA) analysis that refines the student model’s parameters without backpropagation. The training minimizes a combined KL divergence and cross-entropy loss, which improves emotion recognition in terms of both labeled and unlabeled data while maintaining efficiency. 
Jeong et al. \cite{jeong2022multitask} demonstrated the potential of KD in a multitask setting for valence-arousal prediction, facial expression recognition, and action unit prediction. Despite its promise, the use of KD in multitask frameworks for integrating heterogeneous labels remains underexplored.


\subsection{Feature Engineering for MER}

Feature engineering is a key component in Music Emotion Recognition (MER). Earlier works relied heavily on hand-crafted features such as MFCCs, chroma features, and rhythmic descriptors \cite{hizlisoy2021music}, which capture low-level signal properties but fall short in representing higher-level musical semantics \cite{panda2018novel}. Other approaches leverage embeddings from audio models trained on large-scale datasets. For example, VGGish \cite{hershey2017cnn}, a convolutional model trained in a supervised fashion for audio classification, and CLAP \cite{elizalde2023clap}, which is trained with a contrastive audio-text alignment objective, have been used to extract richer representations of audio content.

Pre-trained encoders have also gained traction in MER due to their ability to learn from unlabeled audio data. Among these, MERT \cite{yizhi2023mert} has demonstrated strong performance relative to other pre-trained encoders. For instance, MERT-95M achieves a PR-AUC of 0.134 and an ROC-AUC of 0.764 on the MTG-Jamendo dataset, while the larger MERT-330M model improves these scores slightly to a PR-AUC of 0.14 and an ROC-AUC of 0.765. These results highlight the potential of pre-trained encoders for emotion prediction tasks.

Beyond embeddings, high-level musical features such as key signatures and chord progressions have also been shown to play a significant role in emotion prediction \cite{cho2016music}. In this study, we aim to enhance MER by integrating large-scale audio embeddings with these symbolic, high-level musical descriptors.



\section{Proposed Method}
\label{sec:methods}
In this section, we present our proposed unified multitask learning framework for Music Emotion Recognition (MER), which trains on both categorical and dimensional emotion labels.

The overall framework of our proposed MER system is shown in Figure \ref{fig:framework}. First, we extract three complementary features from audio signals: 1) embeddings from the MERT model \cite{yizhi2023mert}, 2) harmonic representations of chord progressions, and 3) musical key. These features are then concatenated and fed into classifiers in our multitask learning architecture to predict both categorical and dimensional emotion labels simultaneously. Furthermore, we leverage knowledge distillation technique, where teacher models are first trained separately on individual datasets—MTG-Jamendo for categorical labels and DEAM, PMEmo, and EmoMusic for dimensional labels—without multitask-specific branches. These trained teacher models generate soft logits to guide the student model during multitask learning, thereby improving performance across datasets with different label types.

\subsection{Feature Extraction}
To effectively capture music emotion, our proposed framework has two distinct feature extraction pipelines: 1) pre-trained embeddings from MERT and 2) musical features.

\begin{figure}[!t]
\centering
\includegraphics[width=2.1in]{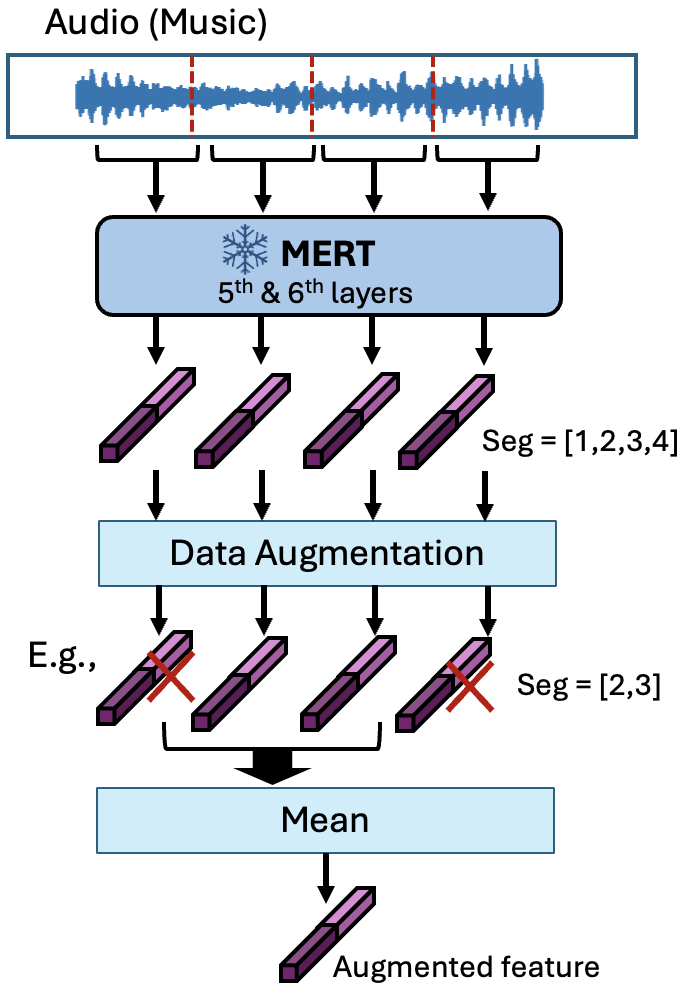}
\caption{MERT feature extraction and data augmentation workflow.}
\label{fig:aug}
\end{figure}

\subsubsection{Pre-trained embeddings (MERT)}
The Music undERstanding model with large-scale self-supervised Training (MERT) \cite{yizhi2023mert} is employed for extracting embeddings from audio files. MERT is able to learn rich musical representations, capturing various musical features that make its embeddings highly effective for downstream tasks such as music emotion recognition (MER). To optimize performance for emotion recognition tasks, we derive embeddings by concatenating the outputs from the 5th and 6th layers of MERT, as these layers have demonstrated superior performance for music emotion prediction through preliminary experimentation.

To improve the model’s robustness, the data augmentation step is incorporated within the MERT feature extraction pipeline as illustrated in Figure \ref{fig:aug}. The data augmentation step is outlined as follows: 

\begin{enumerate} 
\item The audio track is divided into fixed-length segments (i.e., 30 seconds each). 
\item MERT embeddings are obtained for each segment from the 5th and 6th layers.
\item These embeddings from both layers are then combined to create the feature vector for each segment.
\item During training, we randomly select a series of consecutive segments. The starting index and the number of segments are chosen uniformly at random (e.g., Seg = [2,3] in Figure \ref{fig:aug}).
\item The final feature vector is formed by averaging the embeddings of the selected segments.
\end{enumerate}

The intuition behind this augmentation strategy is to expose the model to varying temporal contexts of the same track across training epochs. Emotion in music can vary subtly throughout a piece—different sections might convey slightly different moods due to changes in melody, harmony, rhythm, or instrumentation. By randomly sampling and averaging different combinations of segments, the model learns to generalize across these intra-track variations, rather than overfitting to a single static representation. This not only improves robustness to local variations in musical content but also better reflects the dynamic and evolving nature of emotional expression in music.


\subsubsection{Musical Features (Chord and Key)}
Unlike the segmented and augmented MERT features, musical features (such as chord progressions and key signatures) are calculated over the whole song, representing global harmonic and tonal information. The framework integrates harmonic and tonal information by extracting musical features such as chord progressions and key signatures. A Transformer-based chord recognition model \cite{jonggwon2019bi} is used for extracting the chord progressions of the song. The chord recognition model was evaluated on the following datasets: a subset of 221 songs from Isophonics\footnote{\url{http://isophonics.net/datasets}}(171 songs by the Beatles, 12 songs by Carole King, 20 songs by Queen and 18 songs by Zweieck), Robbie Williams \cite{di2013automatic} (65 songs by Robbie Williams), and a subset of 185 songs from UsPop2002\footnote{\url{https://github.com/tmc323/Chord-Annotations}}. The chord recognition model is evaluated using \texttt{mir\_eval} metrics \cite{raffel2014mir_eval}, achieving Weighted Chord Symbol Recall (WCSR) scores of 83.5\% for Root, 80.8\% for Thirds, 75.9\% for Triads, 71.8\% for Sevenths, 65.5\% for Tetrads, 82.3\% for Maj-min, and 80.8\% for the MIREX categories. These scores are considered acceptable for this study, as most errors are minor, such as misclassifying \textit{A minor} as \textit{C major}.

In our model, each chord is encoded by its root (e.g., \textit{D}) and quality (e.g., \textit{sus4}), representing the chord type, and mapped into an embedding space to capture harmonic patterns. These chord progressions are then converted into MIDI representations based on music theory. For instance, a \textit{C major} chord comprises the notes \textit{C, E, G}, while a \textit{C minor 7th} chord comprises the notes \textit{C, E, G, B}. The start and end times of each chord are mapped to define its duration in the MIDI file. This MIDI representation serves as input to a key detection model.

Key detection is performed using the music21 library \cite{cuthbert2010music21}. To ensure consistency in the representation of chords, we `normalize' them based on the key. Extracted chords from songs in major keys are transposed to C major, while those from minor keys are transposed to A minor. The extracted chord sequences have 13 different chord types such as `major', `minor', `diminished', `augmented', `suspended', and `seventh' chords. The key is encoded based on its mode. For instance, `major' corresponds to C major and `minor' corresponds to A minor. These harmonic features—key (as a scalar) and chords (as sequences of embedded tokens)—are integrated with MERT features and used as input to the model.

\subsubsection{Temporal Modeling of Chord Progressions}
The output of the chord detection model gives us a long sequence of chords. To capture the temporal dependencies and relationships in these harmonic progressions, we model these sequences using a Transformer-based encoder architecture, (marked as `Chord Transformer' in Figure~\ref{fig:framework}). Each chord in the \( t \)-th position of the sequence is represented as a concatenation of its root embedding, \( \mathbf{C}_{\text{root}}^{(t)} \in \mathbb{R}^{d_{\text{root}}} \), and quality embedding, \( \mathbf{C}_{\text{quality}}^{(t)} \in \mathbb{R}^{d_{\text{quality}}} \). The combined embedding for the \( t \)-th chord is expressed as follows:
\begin{equation}
\mathbf{C}^{(t)} = \mathbf{C}_{\text{root}}^{(t)} \oplus \mathbf{C}_{\text{quality}}^{(t)},
\end{equation}
where \( \oplus \) denotes vector concatenation, resulting in \( \mathbf{C}^{(t)} \in \mathbb{R}^{d_{\text{root}} + d_{\text{quality}}} \).

To incorporate the sequential structure of the chord progression, a positional encoding \( \mathbf{P}^{(t)} \in \mathbb{R}^{d_{\text{root}} + d_{\text{quality}}} \) is added to the embeddings:
\begin{equation}
\mathbf{C}_{\text{enc}}^{(t)} = \mathbf{C}^{(t)} + \mathbf{P}^{(t)}.
\end{equation}

The encoded chord embeddings, \( \mathbf{C}_{\text{enc}} = [\mathbf{C}_{\text{enc}}^{(1)}, \mathbf{C}_{\text{enc}}^{(2)}, \ldots, \mathbf{C}_{\text{enc}}^{(T)}] \), represent the sequence of chords for the song, where \( T \) is the total number of chords in the sequence. A special CLS token, \( \mathbf{CLS} \in \mathbb{R}^{d_{\text{root}} + d_{\text{quality}}} \), is prepended to this sequence to aggregate global information:
\begin{equation}
\mathbf{C}_{\text{in}} = [\mathbf{CLS}, \mathbf{C}_{\text{enc}}^{(1)}, \mathbf{C}_{\text{enc}}^{(2)}, \ldots, \mathbf{C}_{\text{enc}}^{(T)}].
\end{equation}

The input sequence is processed by a Transformer encoder with two layers and eight attention heads, generating output embeddings \( \mathbf{C}_{\text{out}} \), where the first token corresponds to the CLS representation:
\begin{equation}
\mathbf{C}_{\text{CLS-out}} = \text{Transformer}(\mathbf{C}_{\text{in}})[0].
\end{equation}

The output \( \mathbf{C}_{\text{CLS-out}} \), taken as the first element of the Transformer output, serves as a global representation of the chord progression. Rather than merely reflecting its positional placement, the CLS token aggregates information from all chord embeddings via the self-attention mechanism.

It is concatenated with the MERT embeddings, \( \mathbf{f}_{\text{MERT}} \), and key embeddings, \( \mathbf{K} \), to form the final combined feature vector:
\begin{equation}
\mathbf{f}_{\text{final}} = \mathbf{C}_{\text{CLS-out}} \oplus \mathbf{f}_{\text{MERT}} \oplus \mathbf{K}.
\end{equation}

The combined feature vector is projected into a latent space using a feedforward layer with 512 units and ReLU activation before being passed to task-specific branches for mood classification and valence-arousal regression. The classification and regression branches each consist of two feedforward layers with 256 hidden units. 

\subsection{Classification with Multitask Learning}
In our proposed multitask learning framework, we have two different branches to handle both categorical and dimensional music emotion prediction tasks. We use a Binary Cross-Entropy (BCE) loss function when training the network on the dataset with categorical labels, such as ``happy'' or ``sad''. To address the class imbalance issue, the BCE loss is weighted based on the frequency of the positive class for each label, which is defined as follows:

\begin{equation}
L_{\text{Cat}}(x, y) = -\frac{1}{c} \sum_{i=1}^{c} \left( w_i y_i \log(x_i) + \bar{w}_i (1 - y_i) \log(1 - x_i) \right),
\end{equation}
where \( x \) is the predicted probabilities for each emotion category, and \( y \) is the corresponding ground-truth binary label (1 for presence and 0 for absence of the category), \( c \) denotes the total number of emotion labels, ensuring that the loss is averaged across all categories, and the term \( w_i \) adjusts the contribution of the positive class, while \( \bar{w}_i \) scales the contribution of the negative class, which can be defined as:
\begin{equation}
w_i = \frac{2}{1 + p_i}, \quad
\bar{w}_i = \frac{2p_i}{1 + p_i},
\end{equation}
where \( p_i \) is the frequency of the positive class for label \( i \).

For dimensional labels, the model predicts continuous valence and arousal (VA) values using a Mean Squared Error (MSE) loss. The MSE loss is given by:

\begin{equation}
L_{\text{VA}}(x, y) = (y_v - x_v)^2 + (y_a - x_a)^2,
\end{equation}
where \( y_v \) and \( y_a \) are the ground-truth valence and arousal values, and \( x_v \) and \( x_a \) are their respective predictions.

We use a selective update strategy to mitigate task interference in our multitask learning framework. 
For the dataset with categorical labels such as MTG-Jamendo, we update only the categorical branch's parameter. Likewise, for the datasets with dimensional labels (e.g., DEAM or PMEmo), we update only the parameters of the dimensional branch. By taking a selective update strategy, the model can effectively learn from heterogeneous datasets with preserved task-specific performance.

\subsection{Knowledge Distillation}
We use the knowledge distillation technique to efficiently train our multitask models by transferring knowledge from the pre-trained teacher models to the multitask student models \cite{hinton2015distilling}. Separate teacher models are trained on categorical and dimensional datasets, each optimized for its respective task. For instance, a teacher model trained on the MTG-Jamendo dataset generates soft labels for categorical labels, while teacher models trained on DEAM, PMEmo, and EmoMusic generate soft labels for dimensional predictions.

The student model learns from both hard labels (ground truth) and soft logits through the Kullback-Leibler (KL) divergence loss, which is defined as:


\begin{equation}
L_{\text{KD}}(s, t) = \sum_{i=1}^{c} t_i \log \left(\frac{t_i}{s_i}\right),
\end{equation}
where \( s \) and \( t \) denote the predicted and teacher-generated soft labels, respectively, \( c \) is the number of output dimensions (i.e., emotion categories for classification, or 2 for valence and arousal in regression), and \( s_i \), \( t_i \) represent the \( i \)-th elements of the student and teacher distributions, respectively.



\subsection{Total Loss Function}
The total loss function combines the task-specific loss with the KL divergence loss derived from the teacher models. Specifically, for categorical and dimensional datasets, the total loss functions are defined as:

\begin{equation}
\label{eq:loss_total_cat}
L_{\text{Total,Cat}} = \alpha \cdot L_{\text{Cat}} + (1 - \alpha) \cdot L_{\text{KD,Cat}},
\end{equation}

\begin{equation}
\label{eq:loss_total_va}
L_{\text{Total,VA}} = \beta \cdot L_{\text{VA}} + (1 - \beta) \cdot L_{\text{KD,VA}},
\end{equation}

where \( \alpha \) and \( \beta \) are hyperparameters that control the balance between the task-specific loss and the KL divergence loss for classification and regression, respectively. \( L_{\text{KD,Cat}} \) and \( L_{\text{KD,VA}} \) refer to the KL divergence losses computed using the soft targets generated by the classification and regression teacher models, respectively. \( L_{\text{Total,Cat}} \) and \( L_{\text{Total,VA}} \) represent the final loss values used to update the classification and regression branches of the multitask model, respectively.




\section{Experimental setup}
\label{sec:exp}
In this section, we discuss the experimental setup of the proposed unified multitask learning framework. 

\subsection{Dataset}
In our experiment, we use four different datasets for either categorical emotion recognition or dimensional emotion prediction tasks. For categorical emotion recognition, we use the \textbf{MTG-Jamendo dataset} \cite{bogdanov2019mtg}, which consists of 18,486 full-length tracks labeled with 56 mood/theme tags such as ``happy'' and ``sad''.

For dimensional emotion prediction, we use three different datasets with valence-arousal (VA) labels which are normalized within the range of 1 to 9:
\begin{itemize}
\item \textbf{DEAM dataset} \cite{aljanaki2017developing}: This dataset contains 1,802 songs from Western pop and rock. It has both dynamic and static VA annotations. We use the static VA annotations. 
\item \textbf{PMEmo dataset} \cite{zhang2018pmemo}: This dataset consists of 794 songs from Western pop. It has both dynamic and static VA annotations. We use the static VA annotations.
\item \textbf{EmoMusic dataset} \cite{soleymani20131000}: This dataset contains 744 songs spanning diverse genres. It has both dynamic and static VA annotations. We use the static VA annotations. 
\end{itemize}

To ensure consistency, we  follow the official train, validation, and test splits for the MTG-Jamendo dataset. For the other datasets which do not provide official training, validation, and test splits, we randomly split the data into 70\% for the training set, 15\% for the validation set, and 15\% for the test set. The details of the datasets used in our experiments are shown in Table \ref{tab:dataset}. 

\begin{table*}[ht!]
\small
\centering
\caption{Details of the dataset.}
\label{tab:dataset}
\begin{tabular}{@{}lccccl@{}}
\toprule
\textbf{Dataset} & \textbf{\# of training data} & \textbf{\# of validation data} & \textbf{\# of test data} & \textbf{Length} & \textbf{Emotion model} \\ 
\midrule
MTG-Jamendo & 9,949 & 3,802 & 4,231 & full & Categoric (56 labels) \\
DEAM & 1,261 & 271 & 270 & 45s & Dimensional \\
PMEmo & 536 & 116 & 115 & full & Dimensional\\
EmoMusic & 495 & 124 & 125 & 45s & Dimensional\\
\bottomrule
\end{tabular}
\end{table*}

\begin{table*}[ht!]
\small
\centering
\caption{Ablation study on influence of input features on performance metrics across datasets. Here, FNN stands for Feedforward Neural Network.}
\label{tab:features_models}
\begin{tabular}{@{}llcccccccc@{}}
\toprule
\textbf{Input feature} & \textbf{Model} & \multicolumn{2}{c}{\textbf{MTG-Jamendo}} & \multicolumn{2}{c}{\textbf{DEAM}} & \multicolumn{2}{c}{\textbf{EmoMusic}} & \multicolumn{2}{c}{\textbf{PMEmo}} \\ 
\cmidrule(lr){3-4} \cmidrule(lr){5-6} \cmidrule(lr){7-8} \cmidrule(lr){9-10}
& & PR-AUC & ROC-AUC & \(R^2_V\) & \(R^2_A\) & \(R^2_V\) & \(R^2_A\) & \(R^2_V\) & \(R^2_A\) \\ 
\midrule
MERT & FNN & 0.1297 & 0.7590 & 0.4972 & 0.5951 & 0.5657 & 0.7404 & 0.5231 & 0.7670 \\

Chord/Key & Transformer + FNN & 0.0531 & 0.6158 & 0.0706 & 0.0770 & 0.0781 & 0.0585 & 0.0149 & 0.0337 \\

MERT + Chord/Key & Transformer + FNN & \textbf{0.1521} & \textbf{0.7806} & \textbf{0.5131} & \textbf{0.6025} & \textbf{0.5957} &  \textbf{0.7489} & \textbf{0.5360} & \textbf{0.7772} \\
\bottomrule
\end{tabular}
\end{table*}

\begin{table*}[ht!]
\small
\centering

\caption{Comparison of performance metrics for our multitask learning model across multiple (test) datasets, when trained on different training sets. Here, J, D, E, and P stand for the MTG-Jamendo, DEAM, EmoMusic, and PMEmo datasets, respectively.}
\label{tab:dataset_fusion}
\begin{tabular}{@{}lccccccccc@{}}
\toprule
\textbf{Training datasets} & \multicolumn{2}{c}{\textbf{MTG-Jamendo (J.)}} & \multicolumn{2}{c}{\textbf{DEAM (D.)}} & \multicolumn{2}{c}{\textbf{EmoMusic (E.)}} & \multicolumn{2}{c}{\textbf{PMEmo (P.)}} \\ 
\cmidrule(lr){2-3} \cmidrule(lr){4-5} \cmidrule(lr){6-7} \cmidrule(lr){8-9}
& PR-AUC & ROC-AUC & \(R^2_V\) & \(R^2_A\) & \(R^2_V\) & \(R^2_A\) & \(R^2_V\) & \(R^2_A\) \\ 
\midrule
Single dataset training (X) & 0.1521 & 0.7806 & 0.5131 & 0.6025 & 0.5957 & 0.7489 & 0.5360 & 0.7772 \\
\midrule
J + D & 0.1526 & 0.7806 & 0.5144 & 0.6046 & - & - & - & - \\
J + E & 0.1540 & 0.7809 & - & - & 0.6091 & 0.7525 & - & - \\
J + P & 0.1522 & 0.7806 & - & - & - & - & 0.5401 & 0.7780 \\
\midrule
J + D + E + P & \textbf{0.1543} & \textbf{0.7810} & \textbf{0.5184} & \textbf{0.6228} & \textbf{0.6512} & \textbf{0.7616} & \textbf{0.5473} & \textbf{0.7940} \\
\bottomrule
\end{tabular}
\end{table*}

\begin{table}[ht!]
\small
\centering
\caption{Comparison of our proposed model with existing models on MTG-Jamendo dataset.}
\label{tab:proposed_model}
\begin{tabular}{@{}lcc@{}}
\toprule
\textbf{Model} & \textbf{PR-AUC} & \textbf{ROC-AUC} \\ 
\midrule
lileonardo \cite{bour2021frequency} & 0.1508 & 0.7747 \\
SELAB-HCMUS \cite{pham2021selab} & 0.1435 & 0.7599 \\
Mirable \cite{tan2021semi} & 0.1356 & 0.7687 \\
UIBK-DBIS \cite{mayerl2021recognizing} & 0.1087 & 0.7046 \\

Hasumi et al. \cite{hasumi2025music}
& 0.0730 & 0.7750\\
Greer et al. \cite{greer2023creating}
& 0.1082 & 0.7354\\
MERT-95M \cite{yizhi2023mert}
& 0.1340 & 0.7640\\
MERT-330M \cite{yizhi2023mert}
& 0.1400 & 0.7650\\

\midrule
Proposed (Ours) & \textbf{0.1543} & \textbf{0.7810} \\
\bottomrule
\end{tabular}
\end{table}

\subsection{Implementation details}
To extract MERT features, the audio tracks are segmented into 30-second clips. We leverage the popular MERT-v1-95M model, accessible via Hugging Face\footnote{\url{https://huggingface.co/m-a-p/MERT-v1-95M}}, to obtain the embeddings. 
Knowledge distillation (KD) is employed, where teacher models pre-trained on each dataset guide the student model during multitask learning. The teacher models are trained separately on each dataset using the same architecture as the student.
For feature extraction, we concatenate the 5th and 6th layer embeddings of MERT and process chord/key features with the Chord Transformer with positional encoding. The hyperparameters for the total loss function (as defined in Equations \ref{eq:loss_total_cat} and \ref{eq:loss_total_va}) are set to \( \alpha = 0.2 \) and \( \beta = 0.2 \). During training, only the relevant loss components are updated based on the dataset type (categorical or dimensional). The student and teacher models are trained for 200 epochs with a batch size of 8, a learning rate of 0.0001, and the Adam optimizer. All models, including both the student and teacher models, are trained on a cluster of four NVIDIA Tesla V100 DGXS GPUs, each with 32 GB of memory. The code is implemented in PyTorch and is available online\footnote{\url{https://github.com/AMAAI-Lab/Music2Emotion}}.


\subsection{Performance Metrics}
Depending on the labeling schemes of datasets, we employ different metrics for evaluating our proposed models. For instance, we use both Precision-Recall AUC (PR-AUC) and Receiver Operating Characteristic AUC (ROC-AUC) for MTG-Jamendo dataset with categorical labels. On the other hand, we use the \(R^2\) scores for valence (\(R^2_V\)) and arousal (\(R^2_A\)) for other three datasets with dimensional labels.

\subsubsection{PR-AUC and ROC-AUC} 
PR-AUC \cite{davis2006relationship} measures precision-recall trade-offs to assess classification performance. This is especially useful for unbalanced datasets. On the other hand, ROC-AUC \cite{davis2006relationship} assesses how well a model performs by analyzing true positive and false positive rates across various decision thresholds. 

\subsubsection{\(R^2\) Scores for Valence and Arousal}
The \(R^2\) score \cite{draper1998applied} quantifies how well the model explains variance in valence and arousal predictions. Higher values indicate better predictive accuracy, with \(R^2_V\) used for valence and \(R^2_A\) for arousal.

\subsection{Baseline Models on MTG-Jamendo}

To evaluate the effectiveness of our approach, we compare it against several representative models developed for the MTG-Jamendo dataset, as shown in Table~\ref{tab:proposed_model}:  
1) \textbf{lileonardo} \cite{bour2021frequency}, a deep CNN that applies frequency-dependent convolutions to mel spectrograms, achieving the top score in the MediaEval 2021 competition;  
2) \textbf{SELAB-HCMUS} \cite{pham2021selab}, which uses a co-teaching training strategy and short audio segments to enhance performance and reduce training time;  
3) \textbf{Mirable} \cite{tan2021semi}, which explores semi-supervised learning through noisy student training and harmonic pitch class profiles (HPCP);  
4) \textbf{UIBK-DBIS} \cite{mayerl2021recognizing}, an ensemble-based model that clusters emotion labels to train specialized classifiers;  
5) \textbf{Hasumi et al.} \cite{hasumi2025music}, which proposes classifier group chains to capture tag dependencies for improved music tagging;  
6) \textbf{Greer et al.} \cite{greer2023creating}, who introduce M3BERT, a multi-task, self-supervised transformer trained via masked reconstruction and fine-tuned for emotion recognition; and  
7) \textbf{MERT} \cite{li2023mert}, a large-scale acoustic music understanding model (with 95M and 330M parameters) pre-trained using pseudo labels from acoustic and musical teachers, demonstrating strong performance across several music understanding tasks.

\section{Results}

In a first experiment, we removed the high-level musical features (i.e., key and chords) from our proposed framework and trained the network using the MERT features only. The performance of the models with these different input feature configurations is shown in Table \ref{tab:features_models}. As can be seen from the table, incorporating both the MERT and high-level musical features significantly enhanced the performance of our model. The model with high-level musical features includes the Chord Transformer model. This achieves the best results, including a PR-AUC of 0.1521 and an ROC-AUC of 0.7806 on the MTG-Jamendo dataset. Furthermore, the framework achieves higher \(R^2_V\) and \(R^2_A\) scores for valence and arousal on the dimensional datasets (DEAM, EmoMusic, and PMEmo), demonstrating the importance of including the musical features.

In a second experiment, we explore how training the network on heterogeneous datasets can improve performance on single datasets. Table \ref{tab:dataset_fusion} shows the comparison of performance metrics when training on multiple datasets. We trained the network on multiple datasets as indicated in the leftmost column. When comparing our results without data fusion to the results for our model  trained on each of the datasets separately (row 1), we notice that the performance clearly increases when adding a second dataset. However, the best performance is reached when training on all datasets (MTG-Jamendo + DEAM + EmoMusic + PMEmo), with a PR-AUC of 0.1543 and an ROC-AUC of 0.7810 on the MTG-Jamendo dataset, as well as the best \(R^2_V\) and \(R^2_A\) scores across the dimensional datasets. These results show the importance of leveraging diverse datasets in a unified multitask framework.

Lastly, we evaluated our proposed model on the MTG-Jamendo dataset using its official train-validation-test split, which enables direct comparison with prior work. As shown in Table \ref{tab:proposed_model}, our model achieves strong performance compared to both earlier baselines such as lileonardo \cite{bour2021frequency} from the MediaEval 2021 competition \cite{tovstogan2021mediaeval}, as well as more recent representative methods \cite{hasumi2025music,greer2023creating, yizhi2023mert}, which reflect current directions in MER modeling. The substantial improvement over these models suggests that our proposed framework is a promising approach for categorical emotion recognition in music. In contrast, other MER datasets do not offer official splits, which makes direct performance comparison less consistent across studies.


These results confirm our proposed framework as a tool to unify categorical and dimensional emotion recognition without sacrificing state-of-the-art performance. The integration of MERT embeddings, musical features, multitask learning, and knowledge distillation proves to be a highly effective method that enhances music emotion recognition on diverse datasets and labeling schemes.

\section{Conclusion}
\label{sec:conclusion}
In this study, we propose a unified multitask learning framework for Music Emotion Recognition (MER) that facilitates training on datasets with both categorical and dimensional emotion labels. Our proposed architecture incorporates knowledge distillation and takes both high-level musical features such as chords and key signatures, as well as pre-trained embeddings from MERT as input, thus enabling it to effectively capture emotional nuances in music. The experimental results demonstrate that our framework outperforms state-of-the-art models on the MTG-Jamendo dataset, including the winners of the MediaEval 2021 competition. Our best model achieves a PR-AUC of 0.1543 and an ROC-AUC of 0.7810. This model is made available open-source online. Our results show that our multitask learning approach enables generalization across diverse datasets, while knowledge distillation facilitates efficient knowledge transfer from teacher to student models. 

In summary, our work provides a robust solution for training on multiple datasets with different types of emotion labels. In future research, we may explore the addition of different musical features as input to the model, refine data augmentation techniques, and expand the framework’s applicability to other affective computing domains.

\section*{Acknowledgment}
This work has received funding from grant no. SUTD SKI 2021\_04\_06 and MOE under grant number MOE-T2EP20124-0014. 
We acknowledge the use of ChatGPT for grammar refinement.

\bibliographystyle{abbrv}
\bibliography{main}

\end{document}